\documentclass[12pt]{article}
\usepackage{epsfig}
\usepackage{latexsym}
\begin{document}
\title{5-dimensional Brans-Dicke Theory and Cosmic Acceleration}
\author{\begin{tabular}{c}\bigskip Li-e Qiang$^{}$\footnote{Email: qqllee815@sohu.com },
Yongge Ma$^{}$\footnote{Email: mayg@bnu.edu.cn }, Muxin
Han$^{}$ \footnote{Email: hamsyncolor@hotmail.com }, Dan Yu, \\
\smallskip Department of Physics, Beijing Normal University,
Beijing, 100875, China
\end{tabular}}
\maketitle
\begin{abstract}
We consider a 5-dimensional scalar-tensor theory which is a direct
generalization of the original 4-dimensional Brans-Dicke theory to
5-dimensions. By assuming that there is a hypersurface-orthogonal
spacelike Killing vector field in the underlying 5-dimensional
spacetime, the theory is reduced to a 4-dimensional theory where
the 4-metric is coupled with two scalar fields. The cosmological
implication of this reduced theory is then studied in the
Robertson-Walker model. It turns out that the two scalar fields
may account naturally for the present accelerated expansion of our
universe. The observational restriction of the reduced
cosmological model is also analyzed.
\end{abstract}
{PACS number(s): 04.50.+h, 98.80.Es}

\section{Introduction}
    The observations on the explosion of type Ia
  supernova indicate that the expansion of our universe might be presently
  accelerating \cite{S,A}. The main component of the universe responsible for
  the acceleration is usually called as
  dark energy, which constitutes about three fourths of the
  whole matter budget of our universe according to the recent WMAP date \cite{Bennett}.
  There are a number of quintessence models of dark energy, which have been put
  forward and most of them involve minimally coupled scalar field
  with different potentials \cite{P,AL,I}. In order to extract a potential suitable
  for describing the universe, two-field quintessence model have also
  been taken account \cite{M,D}. Further models, such as (generalized) chaplygin
  gas \cite{Kam,Bento} and K-essence \cite{Arm,Scherrer}, are proposed
  to account for both dark energy and dark matter in a unified way.
  However, in all these phenomenological models, the scalar
  fields are added by hand and hence their origins are yet
  understood. To explain the accelerated expansion of the universe from
  fundamental physics is now a great challenge. Some researchers
  resort to Brans-Dicke theory \cite{Banerjee,Sen}, which is a
  modified relativistic theory of gravitation apparently compatible
  with Mach's principle \cite{Brans}.  The scalar
  field in Brans-Dicke theory is then expected to account for the desired quintessence
  or K-essence. However, the scalar field can naturally lead to
  cosmological acceleration only when the coupling parameter
  $\omega$ varies with time \cite{Banerjee} or there is a potential
  for it added by hand \cite{Sen}. On the other hand, the models
  of accelerating universe leading by dynamical
  compactification of extra dimensions in Kaluza-Klein theory are
  also proposed \cite{E,F}. As a candidate of fundamental theory, Kaluza-Klein
  theory unifies gravity with electromagnetic field (or Yang-Mill
  field) by certain higher dimensional general relativity
  \cite{Appelguist,Bla}. It is reasonable to propose a fundamental
  theory which combines the advantages of both Brans-Dicke and
  Kaluza-Klein. We thus consider higher dimensional Brans-Dicke theories \cite{KK}
  as available candidates.

  In this paper, to illustrate the physics of these theories we generalize Brans-Dicke
  theory to 5 dimensions and study its effect in the 4-dimensional sensational world.
  On the assumption that there is a hypersurface-orthogonal
  spacelike Killing vector field in the underlying 5-dimensional
  spacetime, the theory is reduced to a 4-dimensional theory where
  the 4-metric is coupled with two scalar fields. Note that if
the extra dimension is compactified as a circle $S^1$ with a
microscopic radius, a Killing vector field may arise naturally in
low energy regime\cite{Bla}. The cosmological implication of this
reduced theory is then studied in the Robertson-Walker model.
Numerical analysis shows that the two scalar fields can be used to
account for the present expansion of our universe, while current
observations impose crucial restrictions on the admissible range
of the free parameters in the reduced model.

\section{Killing reduction of 5-dimensional Brans-Dicke theory}
   Killing reduction of 4-dimensional and 5-dimensional spacetimes have been studied by
   Geroch\cite{R} and Yang et.al.\cite{X}. Let ($M,g_{ab}$) be an $n$-dimensional
   spacetime with a Killing vector field $\xi ^a$, which is everywhere spacelike. Let $S$
denote the collection of all trajectories of $\xi^a$. There is a
one-to-one correspondence between tensor fields
$\hat{T}_{a...c}^{b...d}$ on $S$ and tensor fields
$T_{a...c}^{b...d}$ on $M$ which satisfy
\begin{equation}
\begin{array}{l}
\xi^a T^{b\cdots d}_{a\cdots c}=0,\ \cdots \ ,\xi_d T^{b\cdots
d}_{a\cdots
  c}=0 ,\\
{\mathcal{L}}_\xi T^{b\cdots d}_{a\cdots c}=0 , \label{Li D}
\end{array}
\end{equation}where ${\mathcal{L}}_\xi$ denotes the Lie derivative
with respect to $\xi^a$. Note that the abstract index notation
\cite{wald} is employed. The entire tensor field algebra on $S$ is
completely and uniquely mirrored by tensor field on $M$ subject to
Eq.(\ref{Li D}). Thus, we shall speak of tensor fields being on
$S$ merely as a shorthand way of saying that the fields on $M$
satisfy Eq.(\ref{Li D}). The metric on $S$ is defined as $
h_{ab}=g_{ab}-\lambda ^{-1}\xi _a\xi_b$, where $\lambda \equiv\xi
^a\xi _a$.

The action of our 5-dimensional Brans-Dicke theory is proposed as
\begin{eqnarray}
S_5=\int d^5x\sqrt{-g}[\phi
R^{(5)}-\frac{\omega}{\phi}g^{ab}(\nabla_a\phi)\nabla_b\phi]+\frac{16\pi}{c^4}\int
d^5x\sqrt{-g}L_m^{(5)},
\label{s5}
\end{eqnarray}
where $R^{(5)}$ is the curvature scalar associated with the
5-dimensional space-time metric $g_{ab}$, $\phi$ is a scalar
field, $\omega $ is a coupling constant, and $L_m^{(5)}$
represents the 5-dimensional Lagrangian of the matter fields. The
field equation of $g_{ab}$ derived from $(\ref{s5})$ reads
\begin{eqnarray}
G_{ab}^{(5)}=\frac{8\pi\phi^{-1}}{c^4}T_{ab}^{(5)}
+\frac{\omega}{\phi^2}[(\nabla_a\phi)\nabla_b\phi
-\frac{1}{2}g_{ab}(\nabla^c\phi)\nabla_c\phi]
+\phi^{-1}(\nabla_a\nabla_b\phi-g_{ab}\nabla^c\nabla_c\phi),
\label{g5}
\end{eqnarray}
where $G_{ab}^{(5)}$ is the 5-dimensional Einstein tensor, and
$T_{ab}^{(5)}$ represents the 5-dimensional energy momentum tensor
of matter field with trace $T^{(5)}=T_{ab}^{(5)}g^{ab}$. While the
field equation of $\phi$ is determined by (\ref{s5}) as
\begin{eqnarray}
\nabla^a\nabla_a\phi=\frac{8\pi}{c^4}\frac{T^{(5)}}{4+3\omega}.
\label{phi5}
\end{eqnarray}
We restrict ourselves to 5-dimionsional Brans-Dicke spacetime with
a Killing vector field $\xi^a$, which preserves both $g_{ab}$ and
$\phi$ and is everywhere
 spacelike. Without lose generality we choose a coordinate system
 $\{x^\mu, x^5\},\mu=0,1,2,3,$ adapted to the congruence of
 $\xi^a$, i.e., $
(\frac{\partial}{\partial x^5})^a=\xi^a$. In the case where
$\xi^a$ is not hypersurface-orthogonal, $\xi_\mu\neq0$, $\xi_\mu$
will behave as the electromagnetic 4-potential in the reduced
4-dimensional theory. Since we concern mainly on the effect of the
reduced scalar fields, to simplify the discussion, we will
consider only the case where $\xi^a$ is hypersurface-orthogonal.
Then the line element of $g_{ab}$ reads $ds^2=g_{\mu\nu}dx^\mu
dx^\nu+\lambda dx^5dx^5$. Through Killing reduction, the Ricci
tensor $R^{(4)}_{ab}$ of the 4-metric $h_{ab}$ and the scalar
field $\lambda$ are related to the Ricci tensor $R^{(5)}_{ab}$ of
$g_{ab}$ by
\begin{eqnarray}
R_{ab}^{(4)}=\frac{1}{2}\lambda^{-1}D_aD_b\lambda
-\frac{1}{4}\lambda^{-2}(D_a\lambda)D_b\lambda+h_a^mh_b^nR_{mn}^{(5)}
\label{r4}
\end{eqnarray}
and
\begin{eqnarray}
D^{2}\lambda=\frac{1}{2}\lambda^{-1}(D^a\lambda)D_a\lambda
-2R_{ab}^{(5)}\xi^a\xi^b, \label{lam}
\end{eqnarray}
where $D^2\equiv D^aD_a$ and $D_a$ is the covariant derivative on
$S$ defined by $D_eT_{a\ldots c}^{b\ldots d}=h_e^ph_a^m\ldots
h_c^nh_r^b\ldots h_s^d\nabla_pT_{m\ldots n}^{r\ldots s}$, which
satisfies all the conditions of a derivative operator.

 Let the matter content in Eq.(\ref{s5}) be a 5-dimensional perfect fluid
 with energy momentum tensor
$ T_{ab}^{(5)}=(\rho^{(5)}+P^{(5)})U_a U_b+P^{(5)}g_{ab}$, where
$\rho^{(5)}$ and $P^{(5)}$ are the  energy density and hydrostatic
pressure respectively measured by comoving observers, and $U^a$ is
the 5-velocity of the fluid. Suppose the fluid does not move along
the fifth dimension, then $U_a\xi^a=0$. In the reduced
4-dimensional spacetime $S$, it looks like a 4-dimensional perfect
fluid with energy-momentum: $T_{ab}=(\rho+P)U_a U_b+Ph_{ab}$,
where $\rho\equiv\int\lambda^\frac{1}{2}\rho^{(5)}
dx^5=\lambda^\frac{1}{2}L\rho^{(5)}$ and $P\equiv\int
\lambda^\frac{1}{2}P^{(5)} dx^5=\lambda^\frac{1}{2}LP^{(5)}$; here
$L$ is the coordinate scale of the extra dimension. $T_{ab}^{(5)}$
can then be written into
\begin{eqnarray}
T_{ab}^{(5)}=\frac{1}{L\lambda^{\frac{1}{2}}}[(\rho+P)U_a
U_b+Pg_{ab}].
\end{eqnarray}
Substituting Eq. (\ref{g5}) into (\ref{r4}), we obtain the
4-dimensional field equation for $h_{ab}$ as:
\begin{eqnarray}
G_{ab}^{(4)}&=&\frac{8\pi\phi^{-1}L^{-1}\lambda^{-\frac{1}{2}}}{c^4}T_{ab}^{(4)}
+\frac{1}{2}\lambda^{-1}(D_aD_b\lambda-h_{ab}D^cD_c\lambda)
-\frac{1}{4}\lambda^{-2}[(D_a\lambda)
D_b\lambda-h_{ab}(D^c\lambda) D_c\lambda]\nonumber\\
  &+&\frac{\omega}{\phi^2}[(D_a\phi)
  D_b\phi-\frac{1}{2}h_{ab}(D^c\phi)
D_c\phi]+\phi^{-1}(D_aD_b\phi-h_{ab}D^cD_c\phi)
-\frac{\phi^{-1}}{2}\lambda^{-1}h_{ab}(D^c\lambda) D_c\phi.
\label{g4}
\end{eqnarray}
The reduction of Eq. (\ref{phi5}) leads to the 4-dimensional field
equation of $\phi$ as
\begin{eqnarray}
D^aD_a\phi=-\frac{1}{2}\lambda^{-1}(D^c\lambda) D_c\phi+\frac{8\pi
L^{-1}\lambda^{\frac{1}{2}}}{c^4}(\frac{T^{(4)}+P}{4+3\omega}).
\label{phi4}
\end{eqnarray}
By using Eq. (\ref{lam}), we obtain
\begin{eqnarray}
D^aD_a\lambda=\frac{1}{2}\lambda^{-1}(D^a\lambda)
D_a\lambda-\phi^{-1}(D^a\lambda )D_a\phi+\frac{8\pi
L^{-1}\lambda^{-\frac{1}{2}}\phi^{-1}}{c^4}(\frac{2\omega+2}{4+3\omega}T^{(4)}
-\frac{4\omega+6}{4+3\omega}P). \label{lam4}
\end{eqnarray}
Eqs. (\ref{g4})-(\ref{lam4}) can be regarded as those for a
4-dimensional gravity described by a 4-metric coupled to two
scalar fields, which are equivalent to 5-dimensional Brans-Dicke's
field equations. The two scalar fields appear naturally in the
theory and will be used to explain the present accelerated
expansion of the universe.

\section{Reduced Brans-Dicke cosmology}

To explore the cosmological implication of the reduced Brans-Dicke
theory, we consider the homogeneous and isotropic cosmology with
the 4-dimensional Friedman-Robertson-Walker metric:
\begin{eqnarray}
ds^2=-dt^2+a^2(t)[\frac{dr^2}{1-kr^2}+r^2d\Omega^2],
\end{eqnarray}
where $a(t)$ is the scale factor, and $k$ is the spatial curvature
index. The 4-velocity $U^a$ of the reduced perfect fluid is
required to coincide with that of isotropic observers. Then
Eq.(\ref{g4}) is reduced to
\begin{eqnarray}
-3\frac{\dot{a}}{a}\frac{\dot{\phi}}{\phi}
-\frac{3}{2}\frac{\dot{a}}{a}\frac{\dot{\lambda}}{\lambda}
-\frac{1}{2}\frac{\dot{\lambda}}{\lambda}\frac{\dot{\phi}}{\phi}
+\frac{\omega}{2}(\frac{\dot{\phi}}{\phi})^2
+\frac{8\pi}{c^2}L^{-1}\phi^{-1}\lambda^{-\frac{1}{2}}\rho
=3(\frac{\dot{a}}{a})^2+3\frac{k}{a^2} \label{g00}
\end{eqnarray}
and
\begin{eqnarray}
\frac{\ddot{\phi}}{\phi}+2\frac{\dot{a}}{a}\frac{\dot{\phi}}{\phi}
-\frac{1}{4}({\frac{\dot{\lambda}}{\lambda}})^2
+\frac{1}{2}\frac{\ddot{\lambda}}{\lambda}
+\frac{\dot{a}}{a}\frac{\dot{\lambda}}{\lambda}
+\frac{1}{2}\frac{\dot{\lambda}}{\lambda}\frac{\dot{\phi}}{\phi}
+\frac{\omega}{2}(\frac{\dot{\phi}}{\phi})^2
+\frac{8\pi}{c^2}L^{-1}\phi^{-1}\lambda^{-\frac{1}{2}}P
=-(\frac{2\ddot{a}}{a}+\frac{\dot{a}^2+k}{a^2}), \label{g11}
\end{eqnarray}
where a dot on a letter represents a derivative with respective to
the proper time $t$ of the isotropic observers. While,
Eqs.(\ref{phi4}) and (\ref{lam4}) become the wave equations for
the scalar fields as
\begin{eqnarray}
\ddot{\phi}=-3\frac{\dot{a}}{a}\dot{\phi}
-\frac{1}{2}\frac{\dot{\lambda}}{\lambda}\dot{\phi}
+\frac{8\pi}{c^2}L^{-1}\lambda^{-\frac{1}{2}}\frac{\rho-4P}{3\omega+4}
\label{wphi}
\end{eqnarray}
and
\begin{eqnarray}
\ddot{\lambda}=-3\frac{\dot{a}}{a}\dot{\lambda}
+\frac{1}{2}\frac{(\dot{\lambda})^2}{\lambda}
-\frac{\dot{\phi}}{\phi}\dot{\lambda} +\frac{8\pi
L^{-1}\phi^{-1}\lambda^{\frac{1}{2}}}{c^2}(\frac{2\omega
+2}{3\omega+4}\rho-\frac{2\omega}{3\omega+4}P). \label{wlam}
\end{eqnarray}
The effective energy density and pressure of the scalar fields can
be read out from Eqs.(\ref{g00}) and (\ref{g11}) as
\begin{eqnarray}
\rho_s\equiv\frac{c^2L\lambda^{\frac{1}{2}}}{8\pi}[-3\frac{\dot{a}}{a}\dot{\phi}
-\frac{3}{2}\frac{\dot{a}}{a}\frac{\dot{\lambda}}{\lambda}\phi
-\frac{1}{2}\frac{\dot{\lambda}}{\lambda}\dot{\phi}
+\frac{\omega}{2}\frac{(\dot{\phi})^2}{\phi}] \label{energy}
\end{eqnarray}
and
\begin{eqnarray}
P_s\equiv\frac{c^2L\lambda^{\frac{1}{2}}}{8\pi}[\ddot{\phi}
+2\frac{\dot{a}}{a}\dot{\phi}-\frac{1}{4}({\frac{\dot{\lambda}}{\lambda}})^2\phi
+\frac{1}{2}\frac{\ddot{\lambda}}{\lambda}\phi
+\frac{\dot{a}}{a}\frac{\dot{\lambda}}{\lambda}\phi
+\frac{1}{2}\frac{\dot{\lambda}}{\lambda}\dot{\phi}
+\frac{\omega}{2}\frac{(\dot{\phi})^2}{\phi}]. \label{pressure}
\end{eqnarray}
The scalar fields would play the role of dark energy in the
universe, provided the following conditions are satisfied
\cite{sp}: $ -1.3<\frac{P_s}{\rho_s}<-0.7 $ and $ P_s<0$. From
Eqs.(\ref{energy}) and (\ref{pressure}), one can suppose that
these conditions are very possible to be satisfied in certain
epoch in the evolution of the universe. We now show by numerical
simulation that there are indeed such solutions with accelerated
expansion at the present epoch of the universe. Note that
equations (\ref{g00}) and (\ref{g11}) combined with (\ref{wphi})
and (\ref{wlam}) lead to the matter conservation equation
\begin{eqnarray}
\dot{\rho}+3\frac{\dot{a}}{a}(\rho+P)=0. \label{conserv}
\end{eqnarray}
Since, except for the dark content, the present epoch of the
universe is matter dominated, we put $P=0$ in Eq.(\ref{conserv})
and set $ \rho=\rho _0(a_0/a)^3$, where $\rho_0$ is the current
value of the energy density of visible matter. Thus we suppose
that the dark matter constituent is also involved in the two
scalar fields. From Eqs.(\ref{g00}) and (\ref{g11}) we can obtain
the evolution equation of the Hubble parameter $H(t)$ as
\begin{eqnarray}
\dot{H}=2\frac{\dot{a}}{a}\frac{\dot{\phi}}{\phi}
+\frac{\dot{a}}{a}\frac{\dot{\lambda}}{\lambda}
+\frac{1}{2}\frac{\dot{\lambda}}{\lambda}\frac{\dot{\phi}}{\phi}
-\frac{\omega}{2}(\frac{\dot{\phi}}{\phi})^2
-\frac{8\pi}{c^2}L^{-1}\lambda^{-\frac{1}{2}}\phi^{-1}
\frac{2\omega+3}{3\omega+4}\rho_0(\frac{a_0}{a})^3. \label{htt}
\end{eqnarray}
The wave equations (\ref{wphi}) and (\ref{wlam}) now become
\begin{eqnarray}
\ddot{\phi}=-3\frac{\dot{a}}{a}\dot{\phi}
-\frac{1}{2}\frac{\dot{\lambda}}{\lambda}\dot{\phi}
+\frac{8\pi}{c^2}L^{-1}\lambda^{-\frac{1}{2}}\frac{\rho_0}{3\omega+4}(\frac{a_0}{a})^3
\label{phi}
\end{eqnarray}
and
\begin{eqnarray}
\ddot{\lambda}=-3\frac{\dot{a}}{a}\dot{\lambda}
+\frac{1}{2}\frac{(\dot{\lambda})^2}{\lambda}
-\frac{\dot{\phi}}{\phi}\dot{\lambda}
+\frac{8\pi}{c^2}L^{-1}\phi^{-1}\lambda^{\frac{1}{2}}
\frac{2\omega+2}{3\omega+4}\rho_0 (\frac{a_0}{a})^3.
\label{lambda}
\end{eqnarray}
A necessary and sufficient condition for expansion of the universe
to be accelerating is that the deceleration parameter
$q\equiv-\frac{a\ddot{a}}{\dot{a}^2}$ is negative. Solving
Eqs.(\ref{htt})-(\ref{lambda}) by numerical simulation with
suitable current observed values as initial values, the evolution
of the universe around present epoch can actually be obtained.

The key idea in the dynamical compactification model of
Kaluza-Klein cosmology is that the extra dimensions contract while
the four visible dimensions expand \cite{KK,E,F}. We adopt this
idea to assume that the present universe satisfies
\begin{equation}
a^{3}(t)\lambda^{\frac{n}{2}}(t)=constant,
\end{equation}
where $n$ is a positive number. Then we have
\begin{equation}
\frac{\dot{\lambda}}{\lambda}=-\frac{6}{n}H. \label{dlam}
\end{equation}
Note that the effective 4-dimensional gravitational parameter in
our 5-dimensional theory can be expressed as
$G=(\phi\lambda^{\frac{1}{2}}L)^{-1}$. Hence Eq.(\ref{dlam}) also
implies
\begin{equation}
\frac{\dot{\phi}}{\phi}=\frac{3}{n}H-\frac{\dot{G}}{G}.
\label{dphy}
\end{equation}
Inserting Eqs.(\ref{dlam}), (\ref{dphy}) and $q=-(1+\dot{H}/H^2)$
into Eq.(\ref{htt}), we obtain the relation of $\omega$ and $n$ as
\begin{eqnarray}
\omega=\frac{2(1+q-2z-\varrho)n^2+6zn-18}{(zn-3)^2},\label{omega}
\end{eqnarray}
where $z\equiv\dot{G}/GH$ and $\varrho\equiv\frac{8\pi}{c^2}G
\frac{2\omega+3}{3\omega+4}\rho_0(\frac{a_0}{a})^3H^{-2}$. Taking
account of the current observed values
$\rho_0=(3.8\pm0.2)\times10^{-28}kg\cdot m^{-3}$ and
$H_0=(2.3\pm0.1)\times10^{-18}s^{-1}$ \cite{Freedman}, we can
estimate $\varrho\sim10^{-18}$. Hence the effect of $\varrho$ in
Eq.(\ref{omega}) is so tiny that one can simply neglect it. We
thus expect that the current observational range of $q$ and $z$ as
well as certain physical consideration would constrain the
admissible values of $\omega$ in the 5-dimensional theory and $n$
in the present universe. Taking the current observed values
$q_0=-0.67\pm 0.25$ \cite{Freedman},
$\dot{G}/G=(0.46\pm1.0)\times10^{-12}yr^{-1} \cite{No}$, and the
weak energy condition $\omega \geq-\frac{4}{3}$ \cite{KK},
numerical simulation do give the restricted range of $\omega$ and
$n$ as illustrated in Fig.1. It turns out that the range of $n$ is
restricted to $n\geq2.2$. In general a bigger value of $\omega$
would require a bigger value of $n$. We expect that more
astronomical observations, for example the solar system
experiments, would further constrain the admissible values of
$\omega$ and $n$ \cite{qiang}.

To illustrate the evolutional character of the universe around
present epoch in this model, one can choose some particular values
of $\omega$ and $n$ in the admissible range and obtain desired
evolution curves determined by Eqs.(\ref{htt})-(\ref{lambda}). For
example, one may choose $\omega=-1.2$ and $n=3$ together with
current initial values $H_0=2.3\times10^{-18}s^{-1}$ and
$\rho_0=3.8\times10^{-28}kg\cdot m^{-3}$. Then from
Eqs.(\ref{dlam}) and (\ref{dphy}) one has
$(\frac{\dot{\lambda}}{\lambda})_0=-4.6\times10^{-18}s^{-1}$ and
$(\frac{\dot{\phi}}{\phi})_0=2.3\times10^{-18}s^{-1}$. Let
$a_0=1$, $\lambda(t_0)=1$ and $L=10^{-34}m$, then one gets
$\phi(t_0)=1.5\times10^{44}kg\cdot m^{-4}\cdot s^2$. Taking the
above current values as initial values, the evolution of $a(t)$
with $q(t)$, $\phi(t)$ and $\lambda(t)$ are respectively
illustrated in Fig.2, Fig.3 and Fig.4. Remarkably, from Fig.2 one
can see that the coupling parameter and initial values
 can be chosen in the allowed range such that the current value of the
 deceleration parameter reads $q_0=-0.60$, which coincides with the current
 observation. If the age of the universe today is $t_0=(13\pm1.5)Gyr$ \cite{Freedman},
 the cosmological transition from deceleration to acceleration
 happens at $t=t_0-2.5Gyr$. Fig.3 and Fig.4 show that $\phi(t)$ increases from some
 small value to a bigger constant while $\lambda(t)$ decreases from some large value
 to a smaller constant. So both of them are well behaved. Although our cosmological model is yet to
 be examined, the above result suggests well the scalar
 fields in higher dimensional Brans-Dicke theories to be
 responsible for the present cosmological acceleration.

\section*{Acknowledgement}
This work is  supported in part by NSFC (10205002) and YSRF for
ROCS, SEM. Y. Ma and L. Qiang would also like to acknowledge
support from NSFC (10373003). M. Han would like to acknowledge
support from Undergraduate Research Foundation of BNU. The authors
thank Zhoujian Cao for his valuable help.

\newpage
\begin{figure}[h]
\begin{center}
\includegraphics[width=9cm, bb=140 140 460 460]{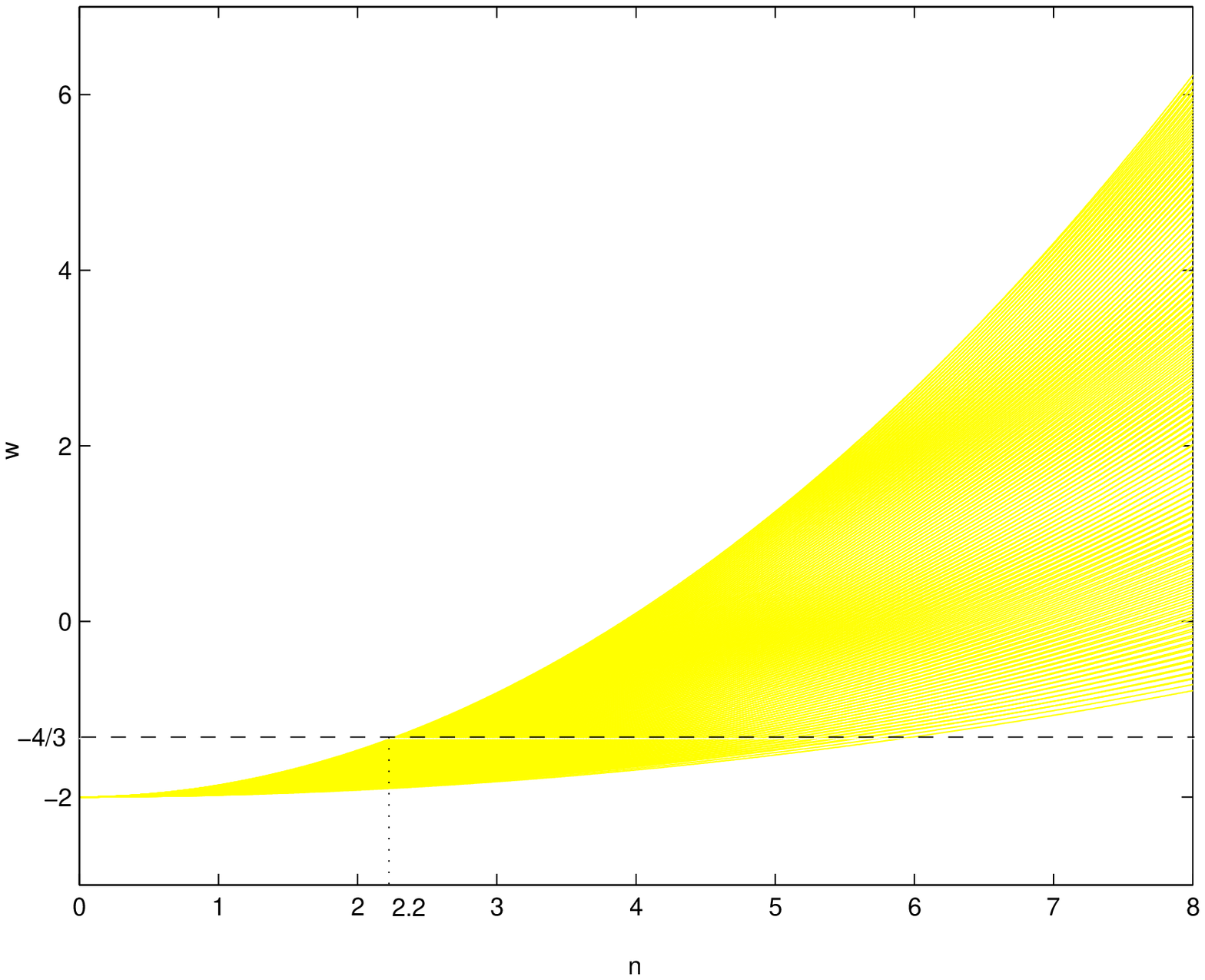}

\caption{The restricted range of $\omega$ and $n$: The current
observation restricts $\omega$ and $n$ in the shadow region, while
the weak energy condition in 5-dimensional Brans-Dicke theory
restricts $\omega\geq-\frac{4}{3}$.} \label{FIG.1}
\end{center}
\end{figure}

\begin{figure}[h]
\begin{center}
\includegraphics[width=9cm, bb=140 140 460 460]{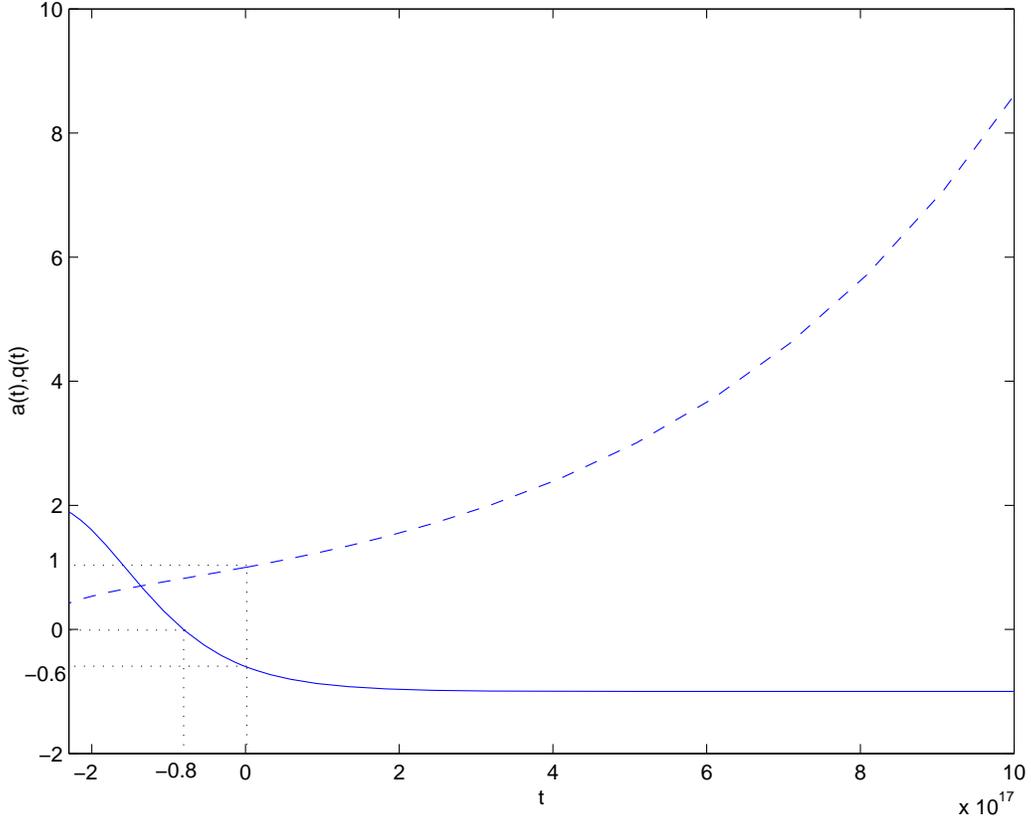}

\caption{The evolution of $a(t)$ (dashed) and the deceleration
parameter $q(t)$ (solid) are shown for the matter-dominated epoch
with $\omega=-1.2$, $n=3$ and the current values: $
\rho_0=3.8\times10^{-28}kg\cdot m^{-3}$,
$H_0=2.3\times10^{-18}s^{-1}$, $a_0=\lambda_0=1$, and
$\phi_0=1.5\times10^{44}kg\cdot m^{-4}\cdot s^2$; here $t=0$
corresponds to today. } \label{FIG.2}
\end{center}
\end{figure}

\begin{figure}[h]
\begin{center}
\includegraphics[width=9cm, bb=140 140 460 460]{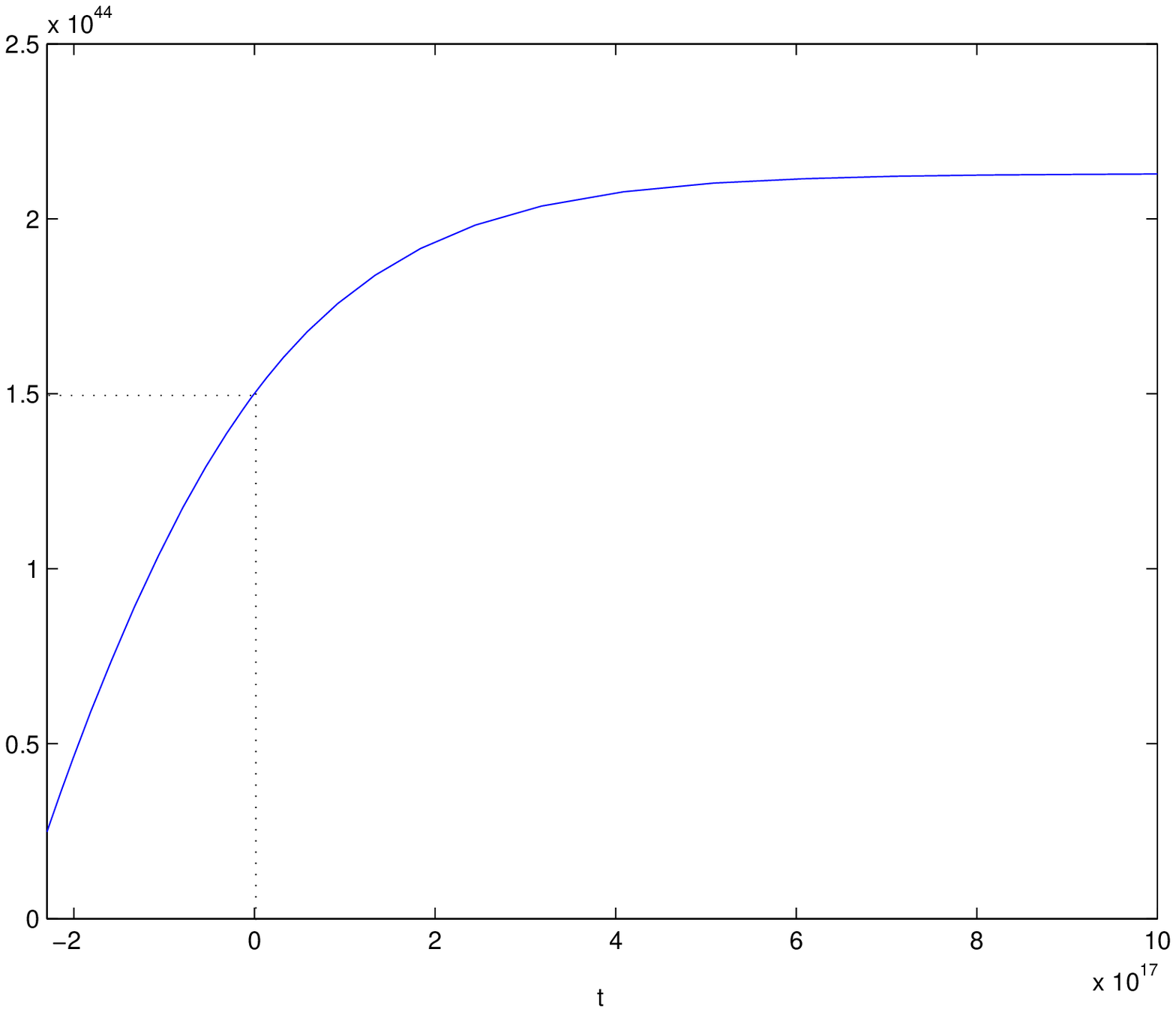}

\caption{The evolution of $\phi(t)$ with the same parameters and
current values as those in Fig.2, and $t=0$ corresponds to today.}
\label{FIG.3}
\end{center}
\end{figure}

\begin{figure}[h]
\begin{center}
\includegraphics[width=9cm, bb=140 140 460 460]{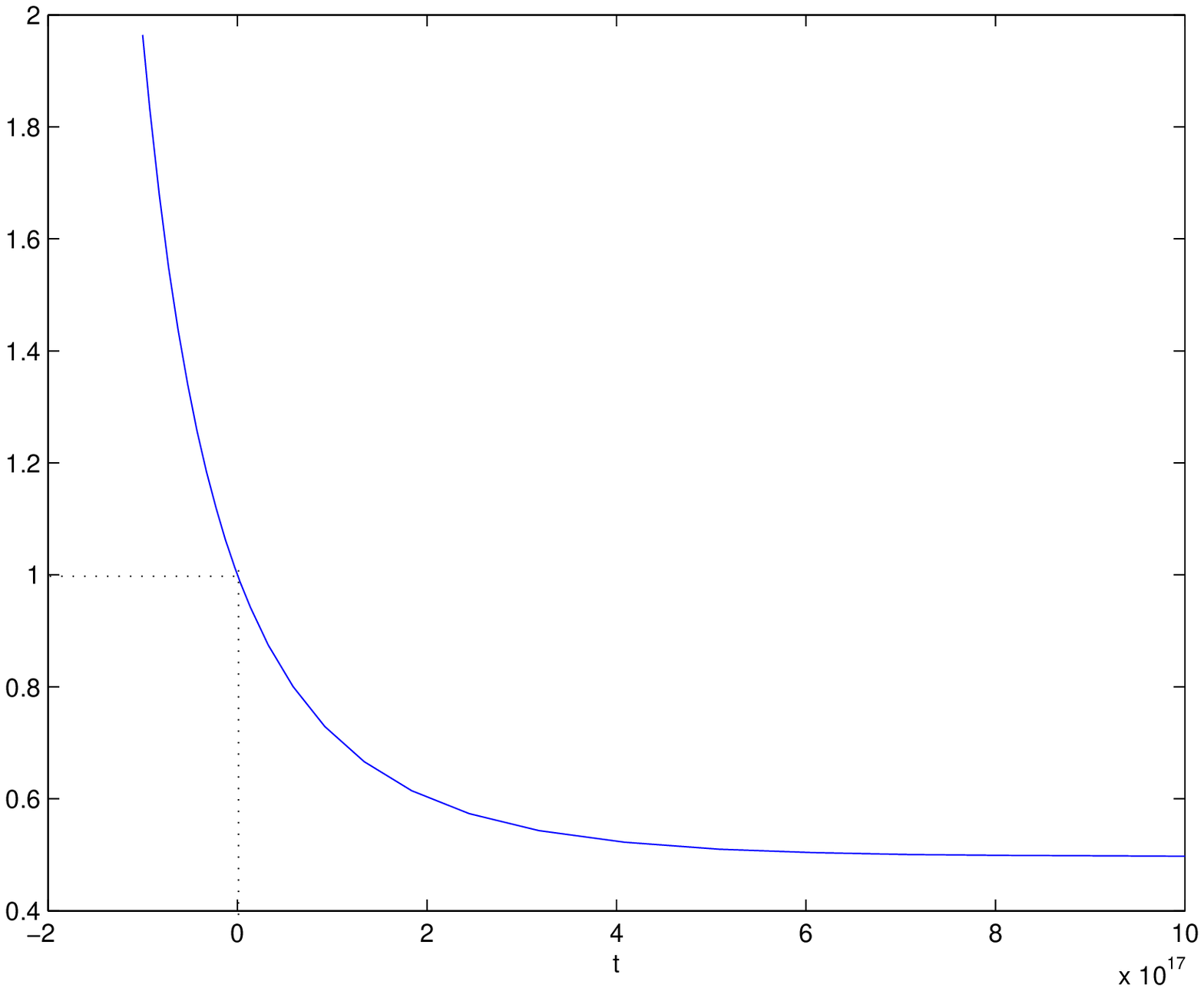}

\caption{The evolution of $\lambda(t)$ with the same parameters
and current values as those in Fig.2, and $t=0$ corresponds to
today. } \label{FIG.4}
\end{center}
\end{figure}

\end{document}